\documentclass[10pt]{ismd08}
\usepackage{graphicx}
\usepackage{cite,./mcite}

\newcommand{\sixoo}{60 \ifb}
\newcommand{\sixooeff}{60 \ifb\,eff$\times2$}
\newcommand{\sixooo}{600 \ifb}
\newcommand{\sixoooeff}{600 \ifb\,eff$\times2$}
\newcommand{\cp}{{\cal CP}}
\newcommand{\MA}{m_A}
\newcommand{\MH}{M_H}
\newcommand{\Mh}{M_h}
\newcommand{\tb}{\tan \beta}

\newcommand{\gsim}
{\;\raisebox{-.3em}{$\stackrel{\displaystyle >}{\sim}$}\;}
\newcommand{\gev}{\,\, \mathrm{GeV}}
\def\citere#1{\mbox{Ref.~\cite{#1}}}
\def\citeres#1{\mbox{Refs.~\cite{#1}}}
\def\si{\sigma}
\def\Ga{\Gamma}
\newcommand{\mev}{\,\, \mathrm{MeV}}
\newcommand{\fh}{{\tt FeynHiggs}}
\newcommand{\HSM}{H^{\SM}}
\newcommand{\SM}{\mathrm {SM}}

\newcommand{\BE}{\begin{equation}}
\newcommand{\EE}{\end{equation}}
\newcommand{\mb}{m_{b}}
\newcommand{\ifb}{$\mbox{fb}^{-1}$}
\newcommand{\Mhmax}{M_h^{\rm max}}

\graphicspath{{figs/}}

\begin{document}

\title{Central Exclusive Production of BSM Higgs bosons at the LHC\protect\footnote{\ \ Based on a talk by V.A. Khoze  at XXXVIII International Symposium on
Multiparticle Dynamics, ISMD08, 15-20 September 2008, DESY Hamburg, Germany.}}
\author{S. Heinemeyer$^1$, V.A. Khoze$^{2}$,
M.G. Ryskin$^{3}$, M. Ta\v{s}evsk\'{y}$^4$ and G. Weiglein$^2$ }

\institute{$^1$Instituto de Fisica de Cantabria (CSIC-UC), Santander, Spain,\\ 
$^2$IPPP, Department of Physics, Durham University, Durham DH1 3LE, U.K.,\\ 
$^3$Petersburg Nuclear Physics Institute, Gatchina, St. Petersburg, 188300, 
Russia,\\ 
$^4$Institute of Physics of the ASCR, 
Na Slovance 2, 18221 Prague 8, Czech Republic}

\maketitle
\begin{abstract}
The prospects for central exclusive diffractive (CED) production of 
MSSM Higgs bosons at the LHC are reviewed.
These processes can
provide important information on the $\cp$-even Higgs bosons,
allowing to probe interesting
regions of the $\MA$--$\tb$ parameter plane. 
The sensitivity of the searches in the forward proton
mode for the Higgs bosons in the so-called CDM-benchmark scenarios
and the effects of fourth-generation models on the CED Higgs production
are briefly discussed.
\end{abstract}

\vspace*{-12cm}
\begin{flushright}
IPPP/08/85 \\
DCPT/08/170
\end{flushright}
\vspace*{10cm}

\section{Introduction}
The physics potential of forward proton
tagging at the LHC has attracted much attention in the
last years, see for instance \cite{INC,DKMOR,Jeff,FP420,pb}. 
The combined detection of both outgoing protons and the centrally
produced system gives access to a unique rich programme
of studies of QCD, electroweak and BSM physics. Importantly,
 these measurements will 
 provide valuable information on the Higgs sector of MSSM and other popular
BSM scenarios, see \cite{KKMRext,HKRSTW,CLP,fghpp}. 

As it is well known,
many models of new physics require an extended
Higgs sector. The most popular extension of the SM 
is the 
MSSM, where the Higgs sector consists of
five physical states. At lowest order the MSSM Higgs
sector is $\cp$-conserving, containing two $\cp$-even
bosons, $h$ and $H$, a  $\cp$-odd boson, $A$, and the charged
bosons $H^\pm$. It can be specified
 in terms of the gauge couplings, the ratio of the two vacuum
expectation values, $\tb \equiv v_2/v_1$, and the mass of the $A$
boson, $\MA$.
The Higgs phenomenology 
in the  MSSM is strongly affected by  higher-order
corrections (see \cite{reviews} for reviews).
Proving that a detected new state is, indeed, a Higgs boson and
distinguishing the Higgs boson(s) of the SM or the MSSM from the states
of other theories will be far from trivial.
In
particular, it will be of utmost importance 
to determine the spin and
$\cp$ properties of a new state and to measure precisely its mass, width
and couplings.

Forward proton detectors
installed at 220~m and 420~m  around ATLAS and / or CMS (see 
\cite{FP420,pb,CMS-Totem}) will provide a rich complementary
physics potential to the  
 ``conventional'' LHC Higgs production channels.
The CED processes are of the form~
~$pp\to p \oplus H \oplus p$, where the $\oplus$ signs denote
 large rapidity gaps on either side of the   centrally produced state.
 If the outgoing protons remain intact and scatter
through small
angles then, to a very good approximation, the primary di-gluon
system obeys a $J_z=0$, $\cp$-even selection rule~\cite{KMRmm}. 
Here $J_z$ is the projection of the total
angular momentum along the proton beam. This 
permits a clean determination of the quantum numbers of the
observed resonance which  will be dominantly produced in a $0^+$ 
state. Furthermore, because the process is exclusive, the
proton  energy
losses are directly related to the central mass,
allowing a potentially excellent mass resolution,
irrespective of the decay channel. The CED
processes allow in principle all the main
Higgs decay modes, $b \bar b$,  $WW$  and $\tau\tau$, to be
observed in the same production channel. In particular, a unique
possibility opens up to study the Higgs Yukawa coupling to bottom
quarks, which, as it is well known, may be difficult
to access in other search channels at the LHC.
Within the MSSM, CED production is even more appealing
than in the SM. 
The coupling of the lightest MSSM Higgs boson to $b \bar b$ and $\tau\tau$ 
can be strongly enhanced for large values of
$\tb$ and relatively small $\MA$. On the other hand, for
larger values of $\MA$ the branching ratio
${\rm BR}(H \to b \bar b)$ 
is much larger
than for a SM Higgs of the same mass. As a consequence, CED 
$H \to b \bar b$ production can be studied in the MSSM up
to much higher masses than in the SM case.

Here we briefly review the analysis of~\cite{HKRSTW} 
where a detailed study of the CED MSSM Higgs
production was performed (see also \citeres{KKMRext,otherMSSM,CLP} for
other MSSM studies). This is updated  by taking into account 
recent theoretical developments in background evaluation~\cite{shuv}
and using an improved version~\cite{FH2.6.2} of the code
\fh~\cite{feynhiggs} employed for the cross section and decay width
calculations. 
These improvements are applied for the CED production of MSSM Higgs
bosons~\cite{HKRSTW} in the 
benchmark scenarios of \cite{benchmark}, the 
so-called CDM-benchmark scenarios, and in a fourth-generation
model.


\section{Signal and background rates and experimental aspects}
\label{sec:backgrounds}

The Higgs signal and background cross sections can be 
approximated  by the simple formulae given in~\cite{KKMRext,HKRSTW}.
For CED production of the MSSM $h, H$-bosons the cross section 
$\si^{\rm excl}$ is 
\begin{equation}
\si^{\rm excl} \, \mbox{BR}^{\rm MSSM} =3 \, {\rm fb} 
  \left(\frac{136}{16+M}\right)^{3.3}
  \left(\frac{120}M\right)^3 
  \frac{\Ga(h/H \to gg)}{0.25\mev} \,\mbox{BR}^{\rm MSSM},
\label{eq1}
\end{equation}
where the gluonic width $\Ga(h/H\to gg)$ and the 
branching ratios for the various MSSM channels,
$\mbox{BR}^{\rm MSSM}$, are calculated with \fh {\tt 2.6.2}~\cite{FH2.6.2}.
The mass $M$ (in GeV) denotes either $\Mh$ or $\MH$.
The normalisation is fixed  at $M=120 \gev$, where
 $\si^{\rm excl} =3$~fb for $\Ga(\HSM \to gg)=0.25 \mev$. 
In \citere{KKMRext,HKRSTW} the uncertainty in the prediction for the CED
cross sections was estimated to be below a factor of $\sim 2.5$. 
According to \cite{DKMOR,krs2,HKRSTW,shuv},
the overall background to the $0^+$ Higgs signal in the
$b \bar b$ mode can be approximated by
\BE
\frac{{\rm d}\si^B}{{\rm d} M} \approx 0.5 \, {\rm fb/GeV} \left[
A\left(\frac{120}{M}\right)^6 +
\frac{1}{2}C \left(\frac{120}{M}\right)^8
                                  \right] 
\label{eq:backbb}
\EE
with $A=0.92$ and $C=C_{\mathrm {NLO}}=0.48 - 0.12 \times (\ln(M/120))$.
This expression holds for a mass window $\Delta M=4-5\gev$ and
summarises several types of
backgrounds: the prolific
$gg^{PP}\to gg$ subprocess can mimic $b\bar b$ production due
to the misidentification of the gluons as $b$ jets;
an admixture of $|J_z|=2$ production;
the radiative $gg^{PP}\to b\bar b g$ background;
due to the non-zero $b$-quark
mass there is also a contribution to the $J_z=0$ 
cross section of order $\mb^2/E_T^2$.
The first term in the square brackets corresponds
to the first three background sources~\cite{HKRSTW}, evaluated 
for $P_{g/b}=1.3 \%$, where $P_{g/b}$ is 
the probability to misidentify a gluon as a $b$-jet for a $b$-tagging
efficiency of 60\%
\footnote{Further improvements in the experimental analysis
could allow to reduce $P_{g/b}$.}%
.~The second term describes the background associated with bottom-mass
terms in the Born amplitude, where one-loop corrections~\cite{shuv}
are accounted for in $C_{\mathrm {NLO}}$. 
The NLO correction suppresses this contribution
by a factor of about 2, or more for larger masses.

The main experimental challenge of running at high luminosity,\
$10^{34} \, {\rm cm}^{-2} \, {\rm s}^{-1}$, 
is the effect of pile-up, which can generate fake signal
events within the
acceptances of the proton detectors
as a result of the coincidence of two or more separate
interactions in the same bunch crossing, 
see \cite{FP420,HKRSTW,CLP,CMS-Totem} for details. 
Fortunately, as established in \cite{CLP}, the pile-up can be brought
under control by using time-of-flight vertexing and  cuts on the number
of charged tracks. Also in the analysis of \cite{HKRSTW}
the event selections and cuts were 
imposed such as to maximally reduce the pile-up background. 
Based on the anticipated improvements for a reduction of the 
overlap backgrounds down to a tolerable level, in the numerical
studies in \cite{HKRSTW,CMS-Totem} and in the new results below the pile-up
effects were assumed to be overcome.

At nominal LHC optics, proton taggers positioned at a distance
$\pm 420$~m from the interaction points of ATLAS and CMS will allow
a coverage of the proton
fractional momentum loss $\xi$ in the range 0.002--0.02,
with an acceptance of around 30\% for a centrally produced system
with a mass around $120 \gev$.
A combination with the foreseen proton detectors at
$\pm 220$~m~\cite{totemRP220} would enlarge
the $\xi$ range up to 0.2.
This would be especially beneficial because of
the increasing acceptance for higher masses \cite{HKRSTW}.
The main selection criteria for $h,H \to b \bar b$ are either two
$b$-tagged jets or two jets with at least one $b$-hadron decaying
into a muon. 
Details on the corresponding selection cuts and triggers for $WW$ and
$\tau\tau$ channels can be found
in \cite{HKRSTW,CMS-Totem,cox1}. Following  \cite{HKRSTW} we
consider four luminosity
scenarios: ``\sixoo'' and ``\sixooo'' refer to running at low and high 
instantaneous luminosity, respectively, using conservative assumptions
for the signal rates and the experimental
sensitivities; possible improvements of both theory and experiment 
 could allow for the scenarios where the
event rates are higher by a factor of 2, denoted as ``\sixooeff'' and
``\sixoooeff''.


\section {Prospective sensitivities  for CED production
of the $\cp$-even Higgs bosons}\label{sec:discovery}

Below we extend the analysis
of the  CED production  of $H \to b \bar b$ and  $H \to \tau \tau $
carried out in \cite{HKRSTW} and consider
the benchmark scenarios of \cite{benchmark}.
The improvements consist of the incorporation of the one-loop
corrections
to the mass-suppressed background~\cite{shuv} and in employing an
updated version of \fh~\cite{feynhiggs,FH2.6.2} for the cross section and
decay width calculations. Furthermore we now also display the limits in
the $\MA$--$\tb$ planes obtained from Higgs-boson searches at the
Tevatron. For the latter we employed  a preliminary version of the
new code {\tt HiggsBounds}, see 
\cite{HiggsBounds} (where also the list of CDF and D0 references for the
incorporated exclusion limits can be found). 
 
\begin{figure}[htb!]
\includegraphics[width=.495\textwidth,height=5cm]{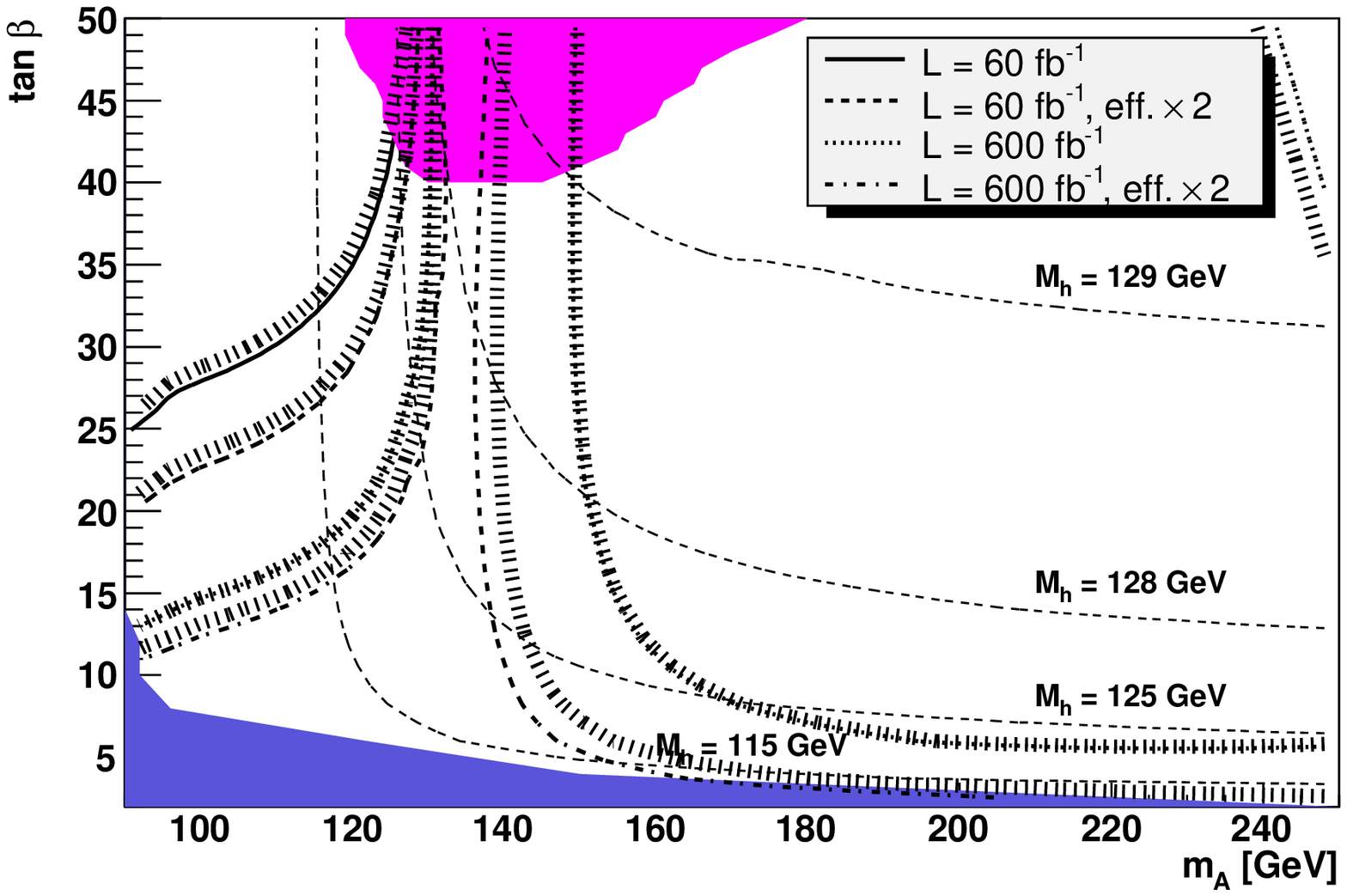}
\includegraphics[width=.495\textwidth,height=5cm]{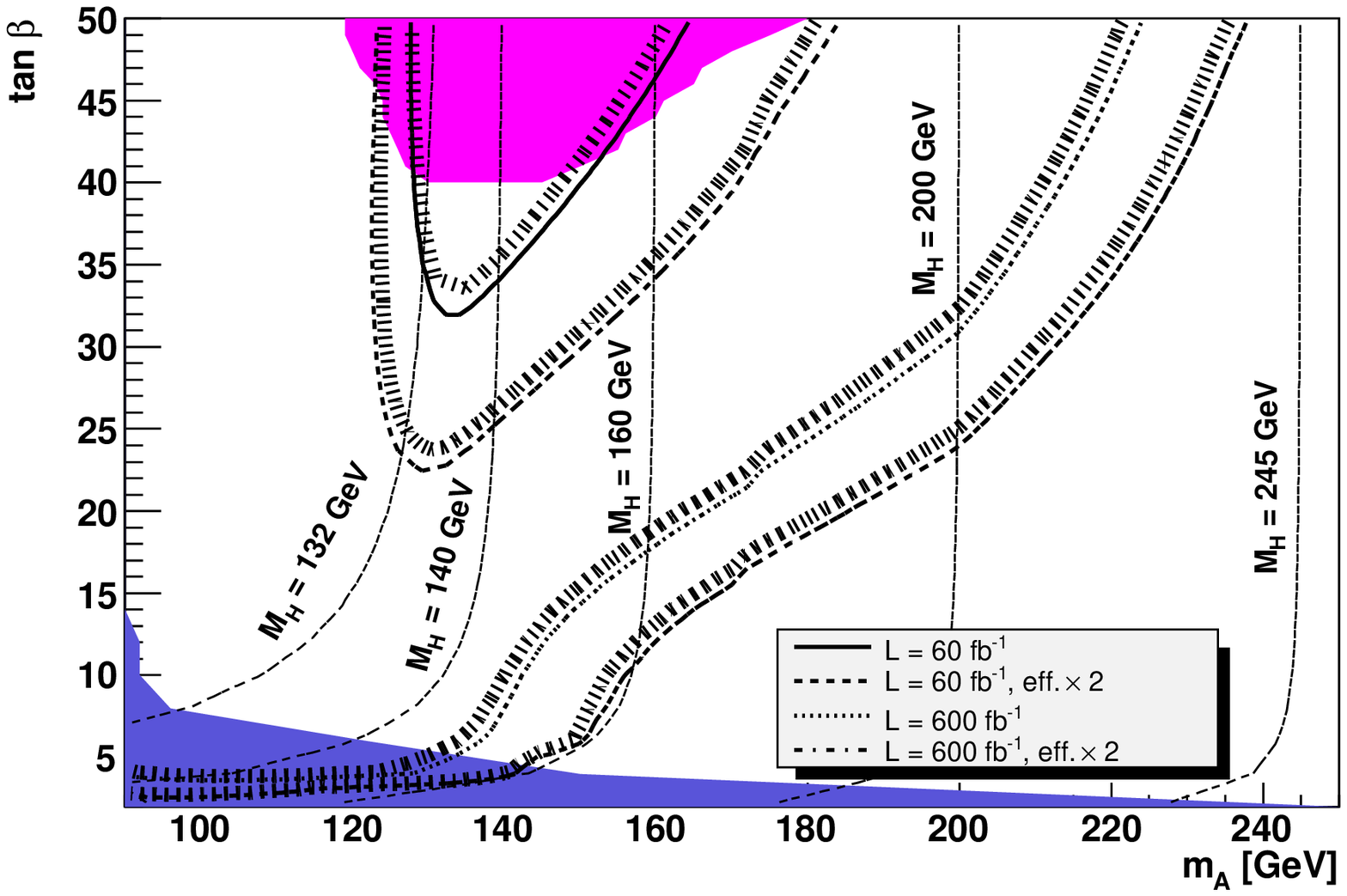}
\vspace{-1.0em}
\caption{Contours of
$3 \si$ statistical significance for the $h \to b \bar b$
channel (left) 
and for the $H \to b \bar b$ channel (right)
in the $\Mhmax$ benchmark scenario with $\mu = +200 \gev$. The results 
were calculated using 
Eqs.~(\ref{eq1}) and (\ref{eq:backbb}) for $A=0.92$ and $C=C_{\mathrm {NLO}}$
for effective luminosities of ``\sixoo'',
``\sixooeff'', ``\sixooo'' and ``\sixoooeff''.
The values of $\Mh$  and $\MH$ are
shown by the contour lines. The medium dark shaded (blue)
regions correspond to the LEP exclusion bounds, while 
the Tevatron limits are  shown by the dark shaded (purple) regions.}
\label{fig:disc}
\vspace{-1em}
\end{figure}

The two plots in Fig.~\ref{fig:disc} exemplify our new
results for the case of the $\Mhmax$ scenario~\cite{benchmark}.
They display the
contours of $3 \si$ statistical significance for the 
$h \to b \bar b$ and $H \to b \bar b$ channels.
The left-hand plot shows that
while the allowed region at high $\tb$ and low $\MA$ can be probed also with 
lower integrated luminosity, in the ``\sixoooeff'' scenario the coverage
at the $3 \si$ level
extends over nearly the whole $\MA$--$\tb$ plane, with the exception of
a  window around $\MA \approx 130-140\gev$ (which widens up for small
values of $\tb$). The coverage
includes the case of a light SM-like Higgs, which corresponds
to the region of large $\MA$. It should be kept in mind that 
besides giving an access to 
the bottom Yukawa coupling, which is a crucial input for determining all
other Higgs couplings~\cite{HcoupLHCSM}, the
forward proton mode would
provide valuable information on the Higgs $\cp$ quantum numbers
and allow a precise Higgs mass measurement and maybe even 
a direct determination of its width.

The properties of the  heavier boson $H$
differ very significantly from the ones of a SM Higgs with the same mass
in the region where $\MH \gsim 150 \gev$. While for a SM Higgs the
$\mbox{BR}(H \to b \bar b)$ is strongly suppressed, the
decay into bottom quarks is the dominant  mode for the MSSM Higgs boson~$H$.
The $3 \si$ significance contours in the $\MA$--$\tb$ plane are
displayed in the right-hand plot of Fig.~\ref{fig:disc}. 
While the area covered in the ``\sixoo'' scenario is to a large extent
already ruled out by Tevatron Higgs searches~\cite{HiggsBounds}, 
in the ``\sixoooeff'' scenario the reach for the
heavier Higgs goes beyond $\MH \approx 235 \gev$ in
the large $\tb$ region. 
At the $5 \si$ level, which is not shown here, the reach
extends up to $\MH \approx 200 \gev$.
Thus, CED production of the $H$ with the subsequent decay to $b \bar b$ 
provides a unique opportunity for accessing its bottom Yukawa
coupling in a mass range where for a SM Higgs boson the $b \bar b$
decay rate  would be negligibly small.
In the ``\sixoooeff'' scenario the discovery of a heavy
$\cp$-even Higgs  with $M_H \approx 140 \gev$ will be possible for all
allowed values of $\tb$. 

In  \cite{CDM} four new MSSM benchmark scenarios were discussed in which
the abundance of the lightest SUSY particle, the lightest neutralino, in
the early universe is compatible within the $\MA$--$\tb$ plane
with the cold dark matter (CDM) constraints as measured by WMAP.
The parameters chosen for the benchmark planes are also in
agreement with electroweak precision and $B$-physics constraints, see
\cite{CDM} for further details.
We studied the prospects of CED Higgs production
for the  $b \bar b$ and 
$\tau\tau$ channels within these so-called CDM benchmark scenarios.
The detailed results will be published elsewhere~\cite{elsewhere}.

\begin{figure}[htb!]
\includegraphics[width=.495\textwidth]{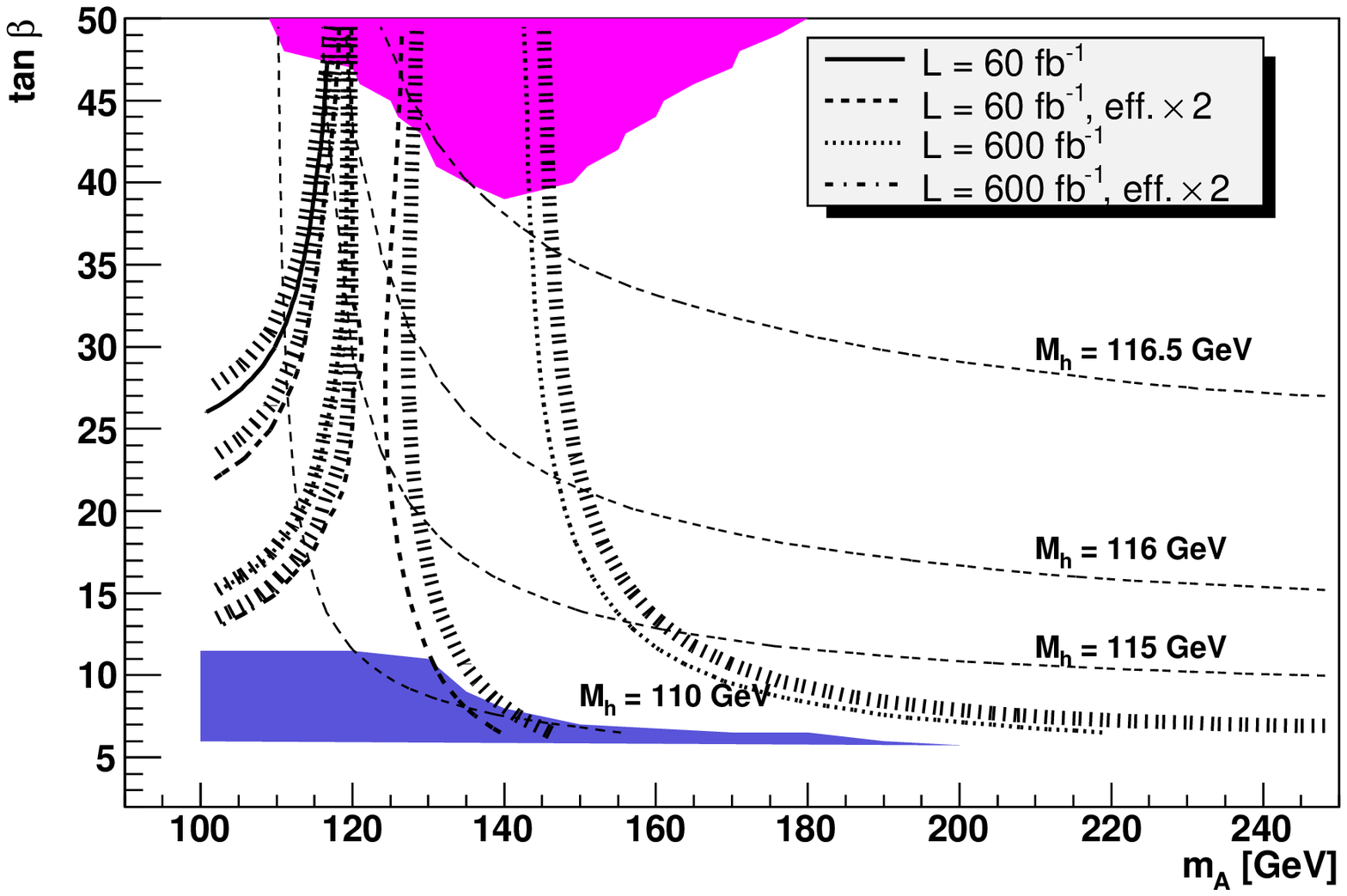}
\includegraphics[width=.495\textwidth]{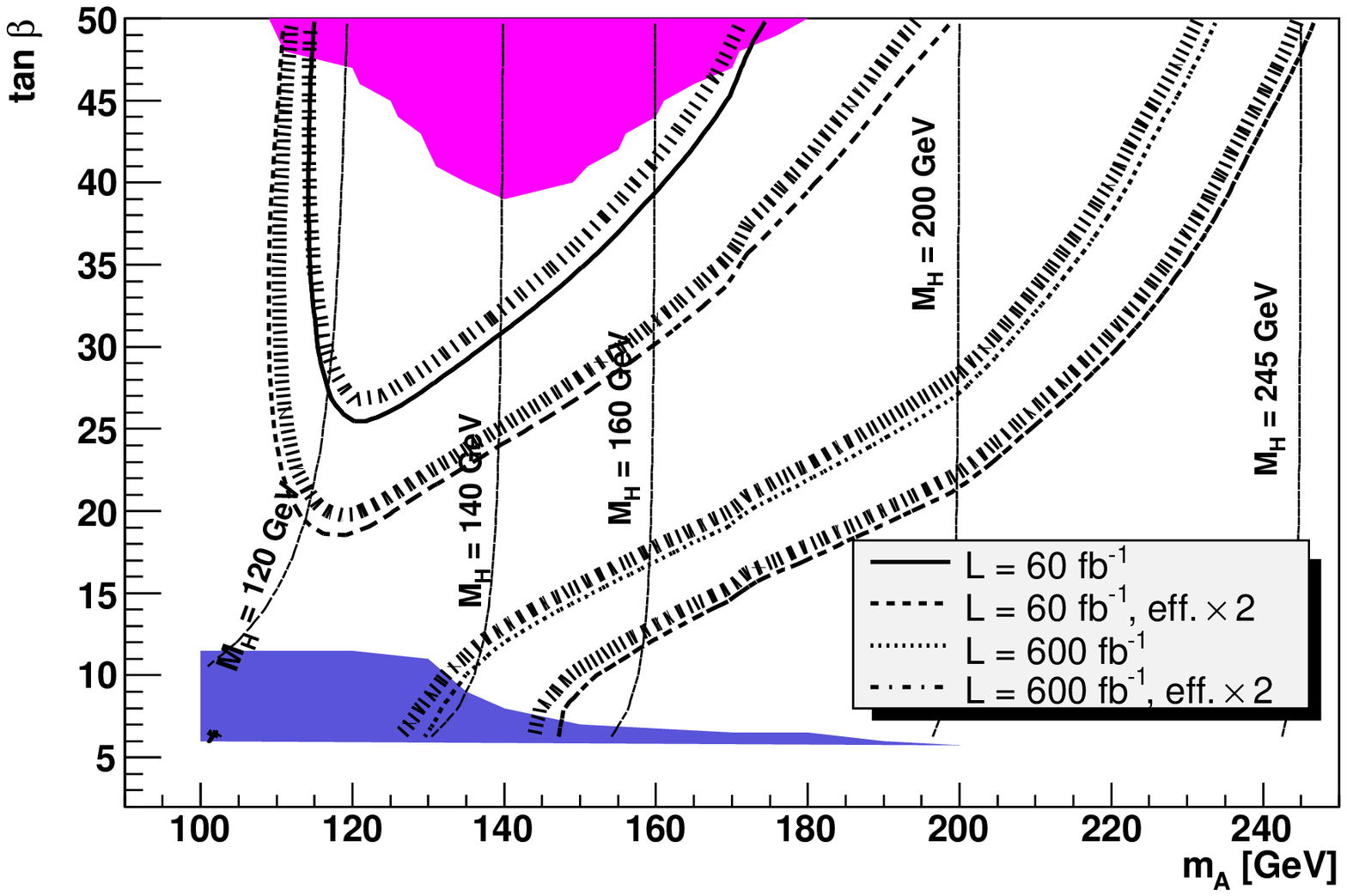}
\vspace{-1em}
\caption{Contours of
$3 \si$ statistical significances
for the $h \to b \bar b$ channel (left) 
and for the $H \to b \bar b$ channel (right)
within the CDM benchmark scenario {\bf P3}. 
The results are calculated 
using the same procedure as in Fig.~\ref{fig:disc}.}
\label{fig:CDM}
\vspace{-1em}
\end{figure}

Here we show two plots in Fig.~\ref{fig:CDM}, exemplifying  our new
results in one of the benchmark planes (called {\bf P3}).
They display the $3 \si$ statistical significances for
the $h \to b \bar b$ and $H \to b \bar b$ processes calculated in the
same way as in the analysis presented in  Fig.~\ref{fig:disc}. 
The results for the $h \to b \bar b$ channel, shown in the left plot of
Fig.~\ref{fig:CDM},  are very similar to the $\Mhmax$ scenario. In the highest
luminosity scenario, ``\sixoooeff'' the $h \to b \bar b$ channel covers
nearly the whole $\MA$--$\tb$ plane, leaving only a small funnel around
$\MA \approx 125 \gev$ uncovered. 
The reach for the $H \to b \bar b$ channel, shown in the right plot of 
Fig.~\ref{fig:CDM}, is slightly better than in the $\Mhmax$
scenario. The area covered in the lowest luminosity scenario, 
``\sixoo'', goes down to $\tb = 25$, so that a larger fraction of the 
parameter space covered at this luminosity is unexcluded by the present 
Tevatron Higgs searches. 
The reach at $\tb = 50$ in the ``\sixoooeff''
scenario goes somewhat beyond $\MH = 240 \gev$ at the $3 \si$ level. 

Finally, we also studied the implications of
a fourth generation of chiral matter on the
CED Higgs production. The interest in this simple kind
of new physics has recently been renewed, see
for example \cite{kribs}.
Within the four-generation scenario the Higgs boson
phenomenology, including the search strategies, is strongly affected.
In particular, the contribution of the fourth-generation quarks gives
rise to an enhancement of the gluonic partial width, $\Ga(H \to gg)$,
by about a factor of 9 compared to the SM case. As a consequence, the 
branching ratios of a light Higgs boson into other final states, such as
$\mbox{BR}(H \to \gamma \gamma)$, are 
significantly suppressed. The CED production rate, on the other hand,
benefits from the enhancement of the gluonic partial width.
The current Tevatron data together with LEP 
limits rule out a Higgs boson in a fourth generation model
below about 210 GeV, apart from a low mass 
window between $115$--$130\gev$.
The CED mechanism offers good prospects to cover this low-mass region
with the rate of the signal $b\bar b$ events exceeding the SM rate by
a factor of about 5--6.
For higher Higgs masses above 210 GeV the rate of the
$H \to WW$  and $H \to ZZ$ events is roughly enhanced by a factor of 9
compared to the SM case. Recall that in this larger mass region
the acceptances of the forward proton detectors (if installed both at 
$\pm 420$~m and $\pm 220$~m from the interaction points) and experimental
selection efficiencies are substantially higher that in the low
mass region \cite{cox1,HKRSTW}. In the  mass range $200$--$250\gev$
the channel $H \to ZZ$ is especially beneficial,
since the only physical background which arises in the semileptonic 
channel and is caused by the $Z$-strahlung process  $pp\to p+Zjj+p$
can be strongly reduced \cite{krs2}.
For illustration we give an estimate of the expected 
number of signal events for the CED Higgs production in a four-generation
case with an integrated luminosity of 
60~fb$^{-1}$. With the proton tagger acceptances and event selection
efficiencies  given in \cite{cox1,HKRSTW} we can expect about 25 
$H \to b \bar b$ events at $\MH = 120 \gev$
and about 45~$WW$ events (when at least one $W$ decays leptonically).
In both cases the evaluated signal-to-background ratio $S/B$ is 
greater than 5.


\subsection*{Acknowledgements}
We are grateful to W.J.~Stirling for collaboration in the early
stages of this work, and we thank O.~Brein, A.~De Roeck, J.~Ellis,
A.~Martin and T.~Tait for useful discussions.
This work was supported in part by the European Community's
Marie-Curie Research Training Network under contract MRTN-CT-2006-035505
`Tools and Precision Calculations for Physics Discoveries at Colliders'
(HEPTOOLS) and MRTN-CT-2006-035657
`Understanding the Electroweak Symmetry
Breaking and the Origin of Mass using the First Data of ATLAS'
(ARTEMIS), and by the project AV0-Z10100502 of the Academy of Sciences of the 
Czech republic and project LC527 of the Ministry of Education of the 
Czech Republic.


\begin{footnotesize}



\end{footnotesize}


\end{document}